\RequirePackage{amsmath} 
\documentclass[runningheads]{llncs}
\usepackage{amsmath}
\usepackage{amsfonts}

\usepackage{graphicx}
\usepackage{bm}
\usepackage[sort]{cite}
\usepackage{algorithm}
\usepackage{algorithmic}
\usepackage{comment}
\usepackage{booktabs}
\usepackage{color}
\usepackage{graphics}

\newtheorem{prp}{Proposition}
\newtheorem{prb}{Problem}
\newtheorem{cor}{Corollary}

\newcommand{\Cost}{\mathrm{Cost}}
\newcommand{\Solve}{\texttt{Solve}}
\renewcommand{\root}{\textrm{root}}
\newcommand{\OPT}{\textrm{OPT}}
\newcommand{\DP}{\texttt{DP}}
\newcommand{\R}{\texttt{R}}
\renewcommand{\S}{\texttt{S}}

\begin{document}
\title{r-Gathering Problems on Spiders: \\ Hardness, FPT Algorithms, and PTASes}
%
%
\author{Soh Kumabe\inst{1,2} \and 
Takanori Maehara\inst{2} 
}
\authorrunning{S. Kumabe and T. Maehara}
%
\institute{The University of Tokyo, Tokyo, Japan \and
RIKEN AIP, Tokyo, Japan \\
\email{soh\_kumabe@mist.i.u-tokyo.ac.jp}\\
\email{takanori.maehara@riken.jp}}
\maketitle              
%

\begin{abstract}
We consider the \emph{min-max $r$-gathering problem} described as follows: 
We are given a set of users and facilities in a metric space.
We open some of the facilities and assign each user to an opened facility such that each facility has at least $r$ users.
The goal is to minimize the maximum distance between the users and the assigned facility.
We also consider the \emph{min-max $r$-gather clustering problem}, which is a special case of the $r$-gathering problem in which the facilities are located everywhere.
In this paper, we study the tractability and the hardness when the underlying metric space is a \emph{spider}, which answers the open question posed by Ahmed et al. [WALCOM'19].
First, we show that the problems are NP-hard even if the underlying space is a spider.
Then, we propose FPT algorithms parameterized by the degree $d$ of the center. 
This improves the previous algorithms because they are parameterized by both $r$ and $d$.
Finally, we propose PTASes to the problems. 
These are best possible because there are no FPTASes unless P=NP.

\end{abstract}


\section{Introduction}

\subsubsection{Background and Motivation.}

We consider the following problem, called the \emph{min-max $r$-gathering problem} ($r$-gathering problem, for short)~\cite{armon2011min}.

\begin{prb}[$r$-gathering problem]
We are given a set $\mathcal{U}$ of $n$ users, a set $\mathcal{F}$ of $m$ facilities on a metric space $(\mathcal{M}, \mathrm{dist})$, and a positive integer $r$.
We open a subset of the facilities and assign each user to an opened facility such that all opened facilities have at least $r$ users.
The objective is to minimize the maximum distance from the users to the assigned facilities.
Formally, the problem is written as follows.
\begin{align} 
    \begin{array}{ll}
    \text{minimize} & \max _{u\in \mathcal{U}} (\mathrm{dist}(u,\pi(u))) \\
    \text{such that} & \pi(u)\in \mathcal{F} \text{ for all } u \in \mathcal{U}, \\
    & |\pi^{-1}(f)|=0 \text{ or } |\pi^{-1}(f)| \ge r \text{ for all } f \in \mathcal{F}.
    \end{array}
\end{align}
\end{prb}
We also consider the \emph{$r$-gather clustering problem}~\cite{akagi2015r,nakano2018simple,ahmed2019r}, which is a variant of the $r$-gathering problem in which the facilities are located everywhere.
\begin{prb}[$r$-gather clustering problem]
We are given a set $\mathcal{U}$ of $n$ users on a metric space $(\mathcal{M}, \mathrm{dist})$ and a positive integer $r$.
We partition $\mathcal{U}$ into arbitrarily many clusters $C_1, \dots,C_k$ such that each user is contained in exactly one cluster, and all clusters contain at least $r$ users.
The objective is to minimize the maximum diameter (distance between the farthest pair) of the clusters.
Formally the problem is written as follows.
\begin{align}
    \begin{array}{ll}
    \text{minimize} & \max \{ \mathrm{diam}(C_1), \dots, \mathrm{diam}(C_k) \} \\
    \text{such that} & \{C_1, \dots, C_k\} \text{ is a partition of } \mathcal{U}, \\
    & |C_1| \ge r, \dots, |C_k| \ge r.
    \end{array}
\end{align}
\end{prb}

These problems have several practical applications, with privacy protection~\cite{sweeney2002k} being a typical one.
Imagine a company that publishes clustered data about their customers.
If there is a tiny cluster, each individual of the cluster can be easily identified.
Thus, to guarantee anonymity, the company requires the clusters to have at least $r$ individuals; this criterion is called the \emph{$r$-anonymity}.
Such clusters are obtained by solving the $r$-gather clustering problem.
Another typical problem is the sport-team formation problem~\cite{ahmed2019r}.
Imagine that a town has $n$ football players and $m$ football courts.
We want to divide the players into several teams, each of which contains at least eleven people and assign a court to each team such that the distance from their homes to the court is minimized.
Such an assignment is obtained by solving the $11$-gathering problem.

In theory, because the $r$-gathering problem and $r$-gather clustering problem are some of the simplest versions of the constrained facility location problems~\cite{drezner2001facility}, several studies have been conducted, and many tractability and intractability results have been obtained so far.
If $\mathcal{M}$ is a general metric space, there is a $3$-approximation algorithm for the $r$-gathering problem, and no algorithm can achieve a better approximation ratio unless P=NP~\cite{armon2011min}.
If the set of locations of the users is a subset of that of the facilities\footnote{This version of the problem is originally called the $r$-gather clustering problem~\cite{aggarwal2010achieving}.}, there is a $2$-approximation algorithm~\cite{aggarwal2010achieving}, and no algorithm can achieve a better approximation ratio unless P=NP~\cite{armon2011min}.
If $\mathcal{M}$ is a line, there are polynomial-time exact algorithms by \emph{dynamic programming} (DP) for the $r$-gathering problem~\cite{akagi2015r,han2016r,nakano2018simple,sarker2019r}, where the fastest algorithm runs in linear time~\cite{sarker2019r}.
The same technique can be implemented to the $r$-gather clustering problem.

If $\mathcal{M}$ is a \emph{spider}, which is a metric space constructed by joining $d$ half-line-shaped metrics together at endpoints\footnote{Ahmed et al.~\cite{ahmed2019r} called this metric space ``star.'' 
In this paper, we followed \url{https://www.graphclasses.org/classes/gc\_536.html}, a part of Information System of Graph Classes and their Inclusions (ISGCI).}
there are \emph{fixed-parameter tractable} (FPT) algorithms parameterized by both $r$ and $d$~\cite{ahmed2019r}
--- More precisely, the running time of their algorithm is $O(n + m + r^d 2^d (r + d) d)$ time\footnote{Ahmed et al~\cite{ahmed2019r}'s original algorithm runs in $O(n + r^2 m + r^d 2^d (r + d) d)$ time, but by combining Sarker~\cite{sarker2019r}'s linear-time algorithm on a line, we obtain this running time.}.
Note that this is not an FPT algorithm parameterized \emph{only by $d$} because it has a factor of $r^d$.
They also posed an open problem that demands a reduction in the complexity of the $r$-gathering problem on a spider.



\subsubsection{Our Contribution}

In this study, we answer the open question that asks the complexity of $r$-gathering problem, posed by Ahmed et al.~\cite{ahmed2019r} by closing the gap between the tractability and intractability of the problems on a spider.

First, we prove that the problems are NP-hard, even on a spider as follows.
\begin{theorem}\label{hardness}
The min-max $r$-gather clustering problem and min-max $r$-gathering problem are NP-hard even if the input is a spider.
\end{theorem}
The proof appears in Section~3 and Appendix~\ref{Ap:hard}.
This implies that some parameterization, such as by the degree $d$ of the center of the spider, is necessary to obtain FPT algorithms for the problems.

Second, we propose FPT algorithms parameterized by $d$ as follows.
\begin{theorem}
\label{tractable}
There is an algorithm to solve the $r$-gather clustering problem on a spider in $O(2^dr^4d^5 + n)$ time, where $d$ is the degree of the center of the spider.
Similarly, there is an algorithm to solve the $r$-gathering problem on a spider in $O(2^dr^4d^5 + n + m)$ time.
\end{theorem}
The proof appears in Section~4.
This result is the best possible in the sense of the number of parameters with superpolynomial dependence because at least one parameter (e.g., degree) is necessary according to Theorem~\ref{hardness}.
Our algorithms have lower parameter dependencies than previous algorithms~\cite{ahmed2019r} because they are parameterized by both $r$ and $d$.
More concretely, our algorithms have no $O(r^d)$ factors.

Finally, we propose \emph{polynomial-time approximation schemes} (PTASes) to the problems.
\begin{theorem}
\label{ptas}
There are PTASes to the $r$-gather clustering problem and $r$-gathering problem on a spider.
\end{theorem}
The proof appears in Section~5.
This result is also the best possible because Theorem~\ref{hardness} implies that there are no \emph{fully polynomial-time approximation schemes} (FPTASes) unless P=NP (Corollary~\ref{nofptas}).
These PTASes can be generalized to the $r$-gather clustering problem and $r$-gathering problem on a tree (see Appendix~\ref{Ap:ptas}).

\section{Preliminaries}


A \emph{spider} $\mathcal{L}=\{l_1,\dots,l_d\}$ is a set of half-lines that share the endpoint $o$ (see Figure~(a)).
Each half-line is called a \emph{leg} and $o$ is the \emph{center}.
The point on leg $l$, whose distance from the center is $x$, is denoted by $(l, x) \in \mathcal{L} \times \mathbb{R}_+$.
It should be noted that $(l, 0)$ is the center for all $l$.
$\mathcal{L}$ induces a metric space whose distance is defined by $\mathrm{dist}((l, x), (l', x')) = |x - x'|$ if $l = l'$ and $x + x'$ if $l \neq l'$.

Let $\mathcal{U}=\{u_1, \dots, u_n\}$ be a set of $n$ users on $\mathcal{L}$. 
A \emph{cluster} $C$ is a subset of users.
The \emph{diameter} of $C$ is the distance between two farthest users in the cluster, i.e., $\mathrm{diam}(C) = \max_{u_i, u_j \in C} \mathrm{dist}(u_i, u_j)$.
 


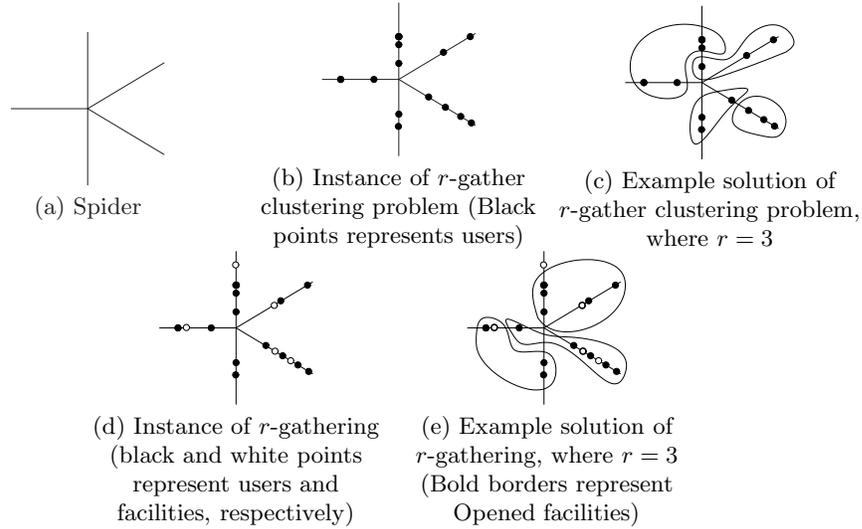
\begin{figure}[tb] 
\centering
\begin{tabular}{c}
    \begin{minipage}{0.33\hsize}
    \centering
          \scalebox{0.4}{
{\unitlength 0.1in%
\begin{picture}(20.0000,20.0000)(0.0000,-20.0000)%
%
\special{pn 13}%
\special{pa 0 1000}%
\special{pa 1000 1000}%
\special{fp}%
\special{pa 1000 0}%
\special{pa 1000 1000}%
\special{fp}%
\special{pa 1000 1000}%
\special{pa 2000 400}%
\special{fp}%
\special{pa 1000 1000}%
\special{pa 2000 1600}%
\special{fp}%
\special{pa 1000 1000}%
\special{pa 1000 2000}%
\special{fp}%
\end{picture}}%
}
          \hspace{1.6cm} (a) Spider
    \end{minipage}
    \begin{minipage}{0.33\hsize}
    \centering
          \scalebox{0.4}{
{\unitlength 0.1in%
\begin{picture}(20.0000,20.0000)(0.0000,-20.0000)%
%
\special{pn 13}%
\special{pa 0 1000}%
\special{pa 1000 1000}%
\special{fp}%
\special{pa 1000 0}%
\special{pa 1000 1000}%
\special{fp}%
\special{pa 1000 1000}%
\special{pa 2000 400}%
\special{fp}%
\special{pa 1000 1000}%
\special{pa 2000 1600}%
\special{fp}%
\special{pa 1000 1000}%
\special{pa 1000 2000}%
\special{fp}%
%
\special{pn 8}%
\special{ar 240 1000 40 40 0.0000000 6.2831853}%
%
\special{sh 1.000}%
\special{ia 240 1000 40 40 0.0000000 6.2831853}%
\special{pn 8}%
\special{ar 240 1000 40 40 0.0000000 6.2831853}%
%
\special{sh 1.000}%
\special{ia 675 1000 40 40 0.0000000 6.2831853}%
\special{pn 8}%
\special{ar 675 1000 40 40 0.0000000 6.2831853}%
%
\special{sh 1.000}%
\special{ia 1000 440 40 40 0.0000000 6.2831853}%
\special{pn 8}%
\special{ar 1000 440 40 40 0.0000000 6.2831853}%
%
\special{pn 8}%
\special{ar 1000 440 40 40 0.0000000 6.2831853}%
%
\special{pn 8}%
\special{ar 1000 440 40 40 0.0000000 6.2831853}%
%
\special{pn 8}%
\special{ar 1000 440 40 40 0.0000000 6.2831853}%
%
\special{sh 1.000}%
\special{ia 1950 1570 40 40 0.0000000 6.2831853}%
\special{pn 8}%
\special{ar 1950 1570 40 40 0.0000000 6.2831853}%
%
\special{sh 1.000}%
\special{ia 1810 1485 40 40 0.0000000 6.2831853}%
\special{pn 8}%
\special{ar 1810 1485 40 40 0.0000000 6.2831853}%
%
\special{sh 1.000}%
\special{ia 1610 1365 40 40 0.0000000 6.2831853}%
\special{pn 8}%
\special{ar 1610 1365 40 40 0.0000000 6.2831853}%
%
\special{sh 1.000}%
\special{ia 1000 790 40 40 0.0000000 6.2831853}%
\special{pn 8}%
\special{ar 1000 790 40 40 0.0000000 6.2831853}%
%
\special{sh 1.000}%
\special{ia 1000 545 40 40 0.0000000 6.2831853}%
\special{pn 8}%
\special{ar 1000 545 40 40 0.0000000 6.2831853}%
%
\special{sh 1.000}%
\special{ia 1585 645 40 40 0.0000000 6.2831853}%
\special{pn 8}%
\special{ar 1585 645 40 40 0.0000000 6.2831853}%
%
\special{sh 1.000}%
\special{ia 1935 440 40 40 0.0000000 6.2831853}%
\special{pn 8}%
\special{ar 1935 440 40 40 0.0000000 6.2831853}%
%
\special{sh 1.000}%
\special{ia 1390 1230 40 40 0.0000000 6.2831853}%
\special{pn 8}%
\special{ar 1390 1230 40 40 0.0000000 6.2831853}%
%
\special{sh 1.000}%
\special{ia 1000 1455 40 40 0.0000000 6.2831853}%
\special{pn 8}%
\special{ar 1000 1455 40 40 0.0000000 6.2831853}%
%
\special{sh 1.000}%
\special{ia 995 1615 40 40 0.0000000 6.2831853}%
\special{pn 8}%
\special{ar 995 1615 40 40 0.0000000 6.2831853}%
\end{picture}}%
}
          \hspace{1.6cm} (b) Instance of $r$-gather clustering problem (Black points represents users)
    \end{minipage}
    \begin{minipage}{0.33\hsize}
    \centering
          \scalebox{0.4}{\input{clusterans.tex}}
          \hspace{1.6cm} (c) Example solution of $r$-gather clustering problem, where $r=3$
    \end{minipage}
\end{tabular}

\begin{tabular}{c}
    \begin{minipage}{0.33\hsize}
    \centering
          \scalebox{0.4}{
{\unitlength 0.1in%
\begin{picture}(20.0000,20.0000)(0.0000,-20.0000)%
%
\special{pn 13}%
\special{pa 0 1000}%
\special{pa 1000 1000}%
\special{fp}%
\special{pa 1000 0}%
\special{pa 1000 1000}%
\special{fp}%
\special{pa 1000 1000}%
\special{pa 2000 400}%
\special{fp}%
\special{pa 1000 1000}%
\special{pa 2000 1600}%
\special{fp}%
\special{pa 1000 1000}%
\special{pa 1000 2000}%
\special{fp}%
%
\special{pn 8}%
\special{ar 240 1000 40 40 0.0000000 6.2831853}%
%
\special{sh 1.000}%
\special{ia 240 1000 40 40 0.0000000 6.2831853}%
\special{pn 8}%
\special{ar 240 1000 40 40 0.0000000 6.2831853}%
%
\special{sh 1.000}%
\special{ia 675 1000 40 40 0.0000000 6.2831853}%
\special{pn 8}%
\special{ar 675 1000 40 40 0.0000000 6.2831853}%
%
\special{sh 1.000}%
\special{ia 1000 440 40 40 0.0000000 6.2831853}%
\special{pn 8}%
\special{ar 1000 440 40 40 0.0000000 6.2831853}%
%
\special{pn 8}%
\special{ar 1000 440 40 40 0.0000000 6.2831853}%
%
\special{pn 8}%
\special{ar 1000 440 40 40 0.0000000 6.2831853}%
%
\special{pn 8}%
\special{ar 1000 440 40 40 0.0000000 6.2831853}%
%
\special{sh 1.000}%
\special{ia 1950 1570 40 40 0.0000000 6.2831853}%
\special{pn 8}%
\special{ar 1950 1570 40 40 0.0000000 6.2831853}%
%
\special{sh 1.000}%
\special{ia 1810 1485 40 40 0.0000000 6.2831853}%
\special{pn 8}%
\special{ar 1810 1485 40 40 0.0000000 6.2831853}%
%
\special{sh 1.000}%
\special{ia 1610 1365 40 40 0.0000000 6.2831853}%
\special{pn 8}%
\special{ar 1610 1365 40 40 0.0000000 6.2831853}%
%
\special{sh 1.000}%
\special{ia 1000 790 40 40 0.0000000 6.2831853}%
\special{pn 8}%
\special{ar 1000 790 40 40 0.0000000 6.2831853}%
%
\special{sh 1.000}%
\special{ia 1000 545 40 40 0.0000000 6.2831853}%
\special{pn 8}%
\special{ar 1000 545 40 40 0.0000000 6.2831853}%
%
\special{sh 1.000}%
\special{ia 1585 645 40 40 0.0000000 6.2831853}%
\special{pn 8}%
\special{ar 1585 645 40 40 0.0000000 6.2831853}%
%
\special{sh 1.000}%
\special{ia 1935 440 40 40 0.0000000 6.2831853}%
\special{pn 8}%
\special{ar 1935 440 40 40 0.0000000 6.2831853}%
%
\special{sh 1.000}%
\special{ia 1390 1230 40 40 0.0000000 6.2831853}%
\special{pn 8}%
\special{ar 1390 1230 40 40 0.0000000 6.2831853}%
%
\special{sh 1.000}%
\special{ia 1000 1455 40 40 0.0000000 6.2831853}%
\special{pn 8}%
\special{ar 1000 1455 40 40 0.0000000 6.2831853}%
%
\special{sh 1.000}%
\special{ia 995 1615 40 40 0.0000000 6.2831853}%
\special{pn 8}%
\special{ar 995 1615 40 40 0.0000000 6.2831853}%
%
\special{sh 0}%
\special{ia 995 175 40 40 0.0000000 6.2831853}%
\special{pn 4}%
\special{ar 995 175 40 40 0.0000000 6.2831853}%
%
\special{sh 0}%
\special{ia 1510 1305 40 40 0.0000000 6.2831853}%
\special{pn 4}%
\special{ar 1510 1305 40 40 0.0000000 6.2831853}%
%
\special{sh 0}%
\special{ia 1500 705 40 40 0.0000000 6.2831853}%
\special{pn 4}%
\special{ar 1500 705 40 40 0.0000000 6.2831853}%
%
\special{sh 0}%
\special{ia 1715 1430 40 40 0.0000000 6.2831853}%
\special{pn 4}%
\special{ar 1715 1430 40 40 0.0000000 6.2831853}%
%
\special{sh 0}%
\special{ia 350 995 40 40 0.0000000 6.2831853}%
\special{pn 4}%
\special{ar 350 995 40 40 0.0000000 6.2831853}%
\end{picture}}%
}
          \hspace{1.6cm} (d) Instance of $r$-gathering (black and white points represent users and facilities, respectively)
    \end{minipage}
    \begin{minipage}{0.33\hsize}
    \centering
          \scalebox{0.4}{\input{r-gatheringans.tex}}
          \hspace{1.6cm} (e) Example solution of $r$-gathering, where $r=3$ (Bold borders represent Opened facilities)
    \end{minipage}
\end{tabular}
\end{figure}

\section{NP-Hardness of r-Gather Clustering on Spider}

We prove Theorem~\ref{hardness} by showing the $r$-gather clustering problem is NP-hard even on a spider.
The NP-hardness of the $r$-gathering problem immediately follows from this result because the $r$-gathering problem is reduced to the $r$-gather clustering problem by putting facilities on the midpoints of the pairs of users.

The strategy for the proof is as follows.
We first introduce the \emph{arrears problem} (Problem~\ref{arrears}) as an intermediate problem.
Then, we reduce the arrears problem to the $r$-gather clustering problem on a spider.
Finally, we prove the strong NP-hardness of the arrears problem.

The arrears problem is the following decision problem.
\begin{prb}[Arrears Problem]
\label{arrears}
We are given $n$ sets $S_1, \dots, S_n$ of pairs of integers, i.e., $S_i = \{ (a_{i,1}, p_{i,1}), \dots, (a_{i,|S_i|}, p_{i,|S_i|}) \}$ for all $i = 1, \dots, n$,
and $m$ pairs of integers $(b_1, q_1), \dots, (b_m, q_m)$.
The task is to decide whether there are $n$ integers $z_1, \dots, z_n$ such that
the following inequality holds for all $j=1, \dots, m$:
\begin{align}
\sum_{a_{i,z_i}\leq b_j}p_{i,z_i}\leq q_j.
\end{align}
\end{prb}
The name of the ``arrears problem'' comes from the following interpretation.
Imagine a person who has pending arrears in his $n$ \emph{payment duties} $S_1, \dots, S_n$.
Each payment duty $S_i$ has multiple options $(a_{i,1}, p_{i,1}), \dots, (a_{i, |S_i|}, p_{i, |S_i|})$ such that he can choose a \emph{payment amount} of \$$p_{i, k}$ with the \emph{payment date} $a_{i, k}$ for some $k$.
Each pair $(b_j, q_j)$ corresponds to his \emph{budget constraint} such that he can pay at most \$$q_j$ until the $b_j$-th day.

The arrears problem itself may be an interesting problem, but here we use this problem as a milestone to prove the hardness of the $r$-gather clustering problem on a spider.
The proof follows the following two propositions.

\begin{prp}[Reduction from the arrears problem]\label{hard1}
If the arrears problem is strongly NP-hard, the min-max $r$-gather clustering problem on a spider is NP-hard.
\end{prp}
\begin{prp}[Hardness of the arrears problem]\label{hard2}
The arrears problem is strongly NP-hard.
\end{prp}

Without loss of generality, we assume that $b_1 < \dots < b_m$ and $q_1 < \dots < q_m$. 
We also assume that $a_{i,1} < \dots < a_{i,|S_i|}$ and $p_{i,1} < \dots < p_{i,|S_i|}$ for all $i = 1, \dots, n$.

\subsection{Reduction from Arrears Problem}

We first prove Proposition~\ref{hard1}.
In this subsection, let $n$ be the number of payment duties and $m$ be the number of budget constraints.

Let $\mathcal{I}$ be an instance of the arrears problem.
We define $L = \max \{ \max_{i} a_{i,|S_i|}, b_m \} +1$ and 
$r = \max \{ \max_{i}p_{i,|S_i|},q_m \}+1$.
We construct an instance $\mathcal{I}'$ of the decision version of the $r$-gather clustering problem on a spider that requires to decide whether there is a way to divide the vertices into clusters each of which has the size of at least $r$ and the diameters of at most $2 L$.

In the construction, we distinguish the legs into two types --- \emph{long} and \emph{short}.
Each long leg corresponds to a payment duty and each short leg corresponds to a budget constraint.
For each payment duty $S_i$, we define a long leg $i$.
We first put $r$ users on $(i, 4L-a_{i,|S_i|}+1)$. Then, we put $p_{i,k+1}-p_{i,k}$ users on $(i,2L-a_{i,k})$ for all $k=1,\dots,|S_i|-1$. 
Finally, we put $r-p_{i,|S_i|}$ users on $(i,2L-a_{i,|S_i|})$.
Each short leg has only one user. The distance from the center to the user is referred to as the \emph{length} of the short leg.
For each $j = 1, \dots, m$, we define $q_j - q_{j-1}$ short legs of length $b_{j-1} + 1$, where we set $q_0 = b_0 = 0$.
We also define $r$ short legs of length $L$.
This construction is done in \emph{pseudo-polynomial} time. 

Now, we prove that $\mathcal{I}'$ has a feasible solution if and only if $\mathcal{I}$ is a YES-instance of the arrears problem.
We first observe a basic structure of clusters in a feasible solution of $\mathcal{I}'$.
The following lemma ensures that the choices of the payment dates on different payment duties are independent of each other.
\begin{lemma}\label{twolegs}
In a feasible solution to $\mathcal{I}'$, there is no cluster that contains users from two different long legs.
\end{lemma}
\begin{proof}
By definition, the distance between the center and a user on a long leg is larger than $L$.
Therefore, the distance between users from two different long legs exceeds $2L$, indicating that they cannot be in the same cluster.
\end{proof}

An \emph{end cluster} of long leg $i$ is a cluster that contains the farthest user of $i$.
The above lemma implies that in a feasible solution, any end cluster of a different long leg is different. 
Intuitively, the ``border'' of the end cluster of long leg $i$ corresponds to the choice from the options of payment duty $S_i$.

\begin{lemma}\label{threesmall}
For each long leg $i$, the following three statements hold.
{\rm (a)} An end cluster of $i$ only contains the users from $i$.
{\rm (b)} There is exactly one end cluster of $i$, and no other cluster consists of only users from leg $i$.
{\rm (c)} Some users on $i$ are not present in the end cluster.
\end{lemma}

\begin{proof}
{\rm (a)} The endpoint of $i$ is distant by more than $2L$ from the center.
{\rm (b)} There are less than $2r$ users on $i$; therefore, they cannot form more then one clusters alone.
{\rm (c)} Users on the point $(i,2L-a_{i,|S_i|})$ are distant from the endpoint of $i$ by more than $2L$; thus they cannot be in the same cluster.
\end{proof}

Lemma~\ref{twolegs} and Statement (c) of Lemma~\ref{threesmall} imply that the users on a long leg who are not contained in end clusters should form a cluster together with users from short legs.
Now, we prove Proposition~\ref{hard1}.

\begin{proof}[Proof of Proposition~\ref{hard1}]
Suppose that we have a feasible solution to the instance of the $r$-gathering problem on a spider that is constructed as mentioned above.
For each long leg $i$, let $u_i$ be the last user that is not contained in end clusters, and $C_i$ be the cluster that contains $u_i$.
Then, the location of $u_i$ is represented as $(i,2L-a_{i,z_i})$ using an integer $z_i$.
We choose the payment date $a_{i,z_i}$ for payment duty $i$.
We prove that these choices of payment dates are a feasible solution to the arrears problem.

As described above, $C_i$ consists of users from leg $i$ and short legs. 
Because there are only $(r-p_{i,|S_i|})+(p_{i,|S_i|}-p_{i,|S_i|-1})+ \dots +(p_{i,z_i+1}-p_{i,z_i})=r-p_{i,z_i}$ users in leg $i$ on the path from the center to the location of $u_i$, $C_i$ should contain at least $p_{i,z_i}$ users on short legs with an at most length of $a_{i,z_i}$.
For the $j$-th budget constraint, by the rule of construction, there are $(q_1-q_0)+ \dots +(q_j-q_{j-1})=q_j$ users on short legs whose length is at most $b_j$.
Suppose $a_{i,z_i}\leq b_j$. We use at least $p_{i,z_i}$ users on short legs whose lengths are at most $a_{i,z_i}\leq b_j$ in the cluster $C_i$. Thus, the sum of $p_{i,z_i}$ among all $i$ with $a_{i,z_i}\leq b_j$ is at most the number of users on short legs whose length is at most $b_j$, that is, $q_j$.
This implies that the budget constraint is valid.

Conversely, suppose that we are given a feasible solution to the instance $\mathcal{I}$ of the arrears problem.
First, for each payment duty $i$ we make a cluster with all users located between $(i,4L-a_{i,|S_i|}+1)$ and $(i, 2L-a_{i,z_i}+1)$, inclusively.
This cluster contains at least $r$ users because there are $r$ users on point $(i,4L-a_{i,|S_i|}+1)$ with a diameter of at most $2L$.
We renumber the payment duties in the non-decreasing order of $a_{i,z_i}$ and proceed them through the order of indices: 
for a payment duty $i = 1, 2, \dots, n$, we make a cluster $C_i$ using all remaining users on leg $i$ and all users from the remaining $p_{i,z_i}$ shortest short legs.
By the construction, these clusters have exactly $r$ users.
We show that the diameter of $C_i$ is at most $2L$.
The diameter is spanned by a long leg and the longest short leg. 
The distance to the long leg in $C_i$ is $2 L - a_{i, z_i}$.
The longest short leg in $C_i$ is the $p_{1,z_1}+ \dots +p_{i,z_i}$-th shortest short leg.
We take the smallest $j$ such that $a_{i,z_i}\leq b_j$.
Then, because the given solution is a feasible solution to $\mathcal{I}$, $p_{1,z_1}+ \dots +p_{i,z_i}\leq q_j$ holds.
Because there are $q_j$ users on short legs with a length of less than $b_{j-1}+1\leq a_{i,z_i}$, the length of the longest short leg in $C_i$ is at most $a_{i,z_i}$.
This gives the diameter of $C_i$ to be at most $2 L$.
Finally, we make a cluster with all remaining users.
Because there are $r$ short legs of length $L$ and all these users are located within the distance $L$ from the center, we can put them into a cluster.
Then, we obtain a feasible solution to $\mathcal{I}'$. 
\end{proof}

\subsection{Strong NP-Hardness of Arrears Problem}

Now we give a proof outline of Proposition~\ref{hard2}; the full proof is given in Appendix~\ref{Ap:hard}.
We reduce the 1-IN-3SAT problem, which is known to be NP-complete~\cite{schaefer1978complexity}.

\begin{prb}[\rm 1-IN-3 SAT problem~\cite{schaefer1978complexity}]
We are given a set of clauses, each of which contains exactly three literals.
Decide whether there is a truth assignment such that all clauses have exactly one true literal.
\end{prb}

\begin{proof}[Proof Outline of Proposition~\ref{hard2}]
Let $n$ and $m$ be the number of boolean variables and clauses, respectively.
For each variable $x_i$, we prepare $N = 3 m (m+2) + 1$ items $T_i$ for a positive literal $x_i$ and $N$ items $\bar{T}_i$ for a negative literal $\bar{x}_i$.
Let $T = \bigcup_i (T_i \cup \bar{T}_i)$ be the set of all items.
Each item $y \in T$ corresponds to a payment duty $\{ (a_{y,1}, p_{y,1}), (a_{y,2}, p_{y,2}) \}$ of two options.
Then, a solution to the arrears problem is specified by a set $X \subseteq T$ of items $y$ such that $a_{y, 2}$ is chosen.
The complement of $X$ is denoted by $\bar{X} = T \setminus X$.
We want to construct a solution to the 1-IN-3SAT problem from a solution $X$ to the arrears problem by 
$x_i = \texttt{true}$ if $y \in X$ for some $y \in T_i$; otherwise $x_i = \texttt{false}$.
We define the payment dates and the amounts suitably to make this construction valid as follows.
%
The payment days consist of two periods: the first period is $\{1, \dots, n\}$ and the second period is $\{n+1, \dots, n+m+2\}$.
For each item $y$, $a_{y,1}$ belongs to the first period and $a_{y,2}$ belongs to the second period. 
Let $i = a_{y,1}$ and $j = a_{y,2}- (n+1)$.
Then, the payment amount $p_{y,1}$ is given in the form of $B^4+\alpha_yB^3+iB^2+i\alpha_yB+j$
where $B$ is a sufficiently large integer, and $\alpha_y$ is a non-negative integer, where $\sum_{y\in T_i}\alpha_y=\sum_{y\in \bar{T}_i}\alpha_y=N$ holds for all $i$.
We define $p_{y,2} = 2p_{y,1}$ for all $y \in T$.

Let $R = (1/2) \sum_{y \in T} p_{y,1} = n N B^4 + n N B^3 + n(n+1)/2 N B^2 + n(n+1)/2 N B + \cdots$. 
We make two budget constraints $(n, R)$ and $(n+m+2, 3R)$.
Then, these constraints hold in equality: Let $x \le R$ be the total payment until $n$. 
Then, the total payment until $n + m + 2$ is $x + 2 (2 R - x) = 4 R - x \le 3 R$.
These inequalities imply that $x = R$. (see Lemma~\ref{halfeq} on Appendix~\ref{Ap:hard}).

We use the first period to ensure that the truth assignment produced by $X$ is well-defined, i.e., if $y \in X$ for some $y \in T_i$, then $y' \in X$ for all $y' \in T_i$.
First, for each $i=1, \dots, n$, we add a budget constraint $(i,iNB^4+iNB^3+(B^3-1))$. By comparing the coefficients of $B^4$ and $B^3$, we have
\begin{align}
\sum_{y\in \bar{X}\cap \bigcup_{j=1}^i(T_j\cup \bar{T}_j)} (B^4 + a_y B^3) \leq i N B^4 + i N B^3.
\end{align}
We can prove that for all $i$, these inequalities hold in equality, i.e., 
\begin{align}
\label{welldefeq}
\sum_{y \in y \in \bar{X}\cap (T_i\cup \bar{T}_i)} (B^4 + a_y B^3) = N B^4 + N B
\end{align}
for all $i$ as follows.
Using the relation between the coefficients of $p_{y,1}$, we have $\sum_{y \in \bar{X}} (i B^2 + i a_y B) \ge \frac{n(n+1)}{2} N B^2 + \frac{n(n+1)}{2} N B$ (see Proposition~\ref{weldef} on Appendix~\ref{Ap:hard}).
Because the budget constraint $(n,R)$ is fulfilled in equality, and the coefficients of $B^2$ and $B$ in $R$ are both $\frac{n(n+1)}{2} N$, this inequality holds in equality, which implies equation \eqref{welldefeq}.
Then, we define the values of $\alpha_y$ appropriately so that only $X \cap (T_i\cap \bar{T}_i) = T_i$ or $X \cap (T_i\cap \bar{T}_i)=\bar{T}_i$ satisfies equation \eqref{welldefeq} (see Proposition~\ref{weldef} on Appendix~\ref{Ap:hard}).
This ensures the well-definedness of the truth assignment.

The second period represents the clauses. Let $Z_i$ be the set of items with $a_{y,2}=i$.
We put a budget constraint $(i,(nN + 2\sum_{j=n+1}^i K_j)B^4+(B^4-1))$ for each $i=n+1, \dots, n+m+2$, where $K_{n+1}, \dots, K_{n+m+2}$ are non-negative integers determined later.
Then, as similar similar to the first period, we can prove that 
\begin{align}
|\bar{X}|+2|X\cap (Z_{n+1}\cup\dots\cup Z_{i})|=nN+2\sum_{j=n+1}^iK_j
\end{align}
for each $i=n+1,\dots,n+m+2$ (see Proposition~\ref{clause} in Appendix B).
This implies that $|X\cap Z_i|=K_i$ for each $i = n+1, \dots, n+m+2$.
The budget constraint on day $i \ge n + 3$ corresponds to the $i-(n+2)$-th clause.
For $i=n+3, \dots, n+m+2$, we set $K_i=1$.
Then, we have $|X\cap Z_i|=1$, i.e., exactly one literal in the $i-(n+2)$-th clause is $\mathtt{true}$. 
The budget constraints on day $n+1$ and $n+2$ are used for the adjustment.
Because $\{ Z_{n+1}, \dots, Z_{n+m+2} \}$ forms a partition of items, we have $|X \cap Z_{n+1}|+|X \cap Z_{n+2}|=|X|-(|X\cap Z_{n+3}| + \dots + |X\cap Z_{n+m+2}|) = N-m$.
Moreover, because the constant term $e_{y,0}$ of $p_{y,1}$ is $e_{y,0} = i-(n+1)$ for all $y\in Z_i$ and $i=n+1,\dots,n+m+2$, we have $\sum_{y\in X} e_{y,0} = \sum_{i=n+1}^{n+m+2}(i-(n+1))|X\cap Z_i|$.
By solving these equations, we obtain $K_{n+1} = |X \cap Z_{n+1}|$ and $K_{n+1} = |X \cap Z_{n+2}|$. 
Because all values appearing in $\mathcal{I}'$ are at most $2B^4$, we can take $B$ in a polynomial of $n,m$.
Thus, the hardness proof is completed.
\end{proof}

The following is a consequence of the construction.
\begin{cor}
\label{nofptas}
The $r$-gather clustering problem on a spider does not admit an FPTAS unless P=NP.
\end{cor}
\begin{proof}
The diameter of the constructed spider is bounded by $O(n + m)$. 
Let us take such an instance.
If there is an FPTAS for the $r$-gathering problem on a spider, 
by taking $\epsilon = 1 / (c(n + m))$ for a sufficiently large constant $c$, we get an optimal solution because the optimal value is an integer at most $O(n+m)$.
This contradicts the hardness.
\end{proof}

\section{FPT Algorithm for r-Gather Clustering and r-Gathering on Spider}

We prove Theorem~\ref{tractable} by obtaining FPT algorithms to solve the $r$-gather clustering problem and $r$-gathering problem on a spider parameterized by the number $d$ of legs.
Due to the space limitation, we put all the pseudocodes in Appendix~\ref{ap:pseudocode}. 

First, we exploit the structure of optimal solutions.
After that, we give a brute-force algorithm. Finally, we accelerate it by DP.

We denote the coordinate of user $u$ by $(l(u),x(u))$.
Without loss of generality, we assume that $x(u_1) \le \dots \le x(u_n)$. 
We use this order to explain a set of users; for example, ``the first (resp. last) $k$ users on leg $l$'' indicates the users with $k$ smallest (resp. largest) index among all users on leg $l$.
We choose an arbitrary leg and consider all the users on the center as being located on this leg.

We introduce a basic lemma about the structure of a solution.
A cluster is \emph{single-leg} if it contains users from a single leg; otherwise, it is \emph{multi-leg}.
Ahmed et al.~\cite{ahmed2019r} showed that there is an optimal solution that has a specific single-leg/multi-leg structure as follows.

\begin{lemma}[{\rm \cite[Lemma~2]{ahmed2019r}}]
For both $r$-gather clustering problem and $r$-gathering problem, there is an optimal solution such that for all leg $l$, some users from the beginning (with respect to the order described above) are contained in multi-leg clusters, and the rest of them are contained in single-leg clusters.
\end{lemma}

Now, we concentrate on the structure of multi-leg clusters.
Let $C$ be a multi-leg cluster.
Let $u_i$ be the last user in $C$ and $u_j$ be the last user with $l(u_i)\neq l(u_j)$ in $C$.
A \emph{ball part} of $C$ is the set of users in $C$ whose indices are at most $j$ and a \emph{segment part} of $C$ is the set of the remaining users in $C$. 
$C$ is \emph{special} if $C$ contains all users on $l(u_i)$ and the ball part is $\{u_1, \dots, u_k\}$ for some integer $k$.
The list of multi-leg clusters $\{C_1, \dots ,C_t\}$ are {\it suffix-special} if for all $1\leq i\leq t$, $C_i$ is a special when we only consider the users in $C_i, \dots ,C_t$.

The following lemma is the key to our algorithm.
We omit the proof because it is a reformulation of Lemma~3 and Lemma~8 in \cite{ahmed2019r} using Lemma~2 in \cite{nakano2018simple}. 
\begin{lemma}[{\rm Reformulation of \cite[Lemmas~3 and 8]{ahmed2019r} by \cite[Lemma~2]{nakano2018simple}}]
Suppose $|\mathcal{U}|\geq 1$ and there exists an optimal solution without any single-leg cluster.
Then, there is an optimal solution such that all clusters contain at most $2r-1$ users, 
and there exists a special cluster.
%
\end{lemma}

By definition, the segment part of a cluster is non-empty and contains users from a single leg.
By removing a special cluster and applying the lemma repeatedly, we can state that there is an optimal solution consisting of a suffix-special family of multi-leg clusters.

Algorithm~\ref{subalg} in Appendix~\ref{ap:pseudocode} is a brute-force algorithm that enumerates all suffix-special families of multi-leg clusters.
The correctness is clear from the definition.
For each enumerated clusters, we fix them and consider the remaining problem, which consists of single-leg clusters.
Thus, the optimal solution is obtained by solving the line case problems independently for each leg.

Now, we accelerate this algorithm by DP.
We observe that instead of remembering all data of $C$, it is sufficient to remember (1) the size of $C$ (to avoid creating too-small clusters) and (2) the index of the last user in the ball part of $C$ (to calculate the diameter/cost of the cluster).
Here, (2) implies that if we know the last user $u$ in the ball part of $C$ and last user $v$ in the segment part of $C$, the diameter/cost of $C$ is computed because $C$ is spanned by $u$ and $v$. 
Below, we denote the diameter/cost of the multi-leg cluster by $\Cost(v,u)$ for both problems.

We also accelerate the process for single-leg clusters. 
As pre-processing for all leg $l$ and all integers $k$ from $0$ to the number of users on leg $l$, we first compute the optimal value of the problem that only considers the last $k$ users on leg $l$.
For each user $u_i$, we denote the optimal objective value for the set of users on leg $l(u_i)$ whose indices are greater than $i$ and no less than $i$ by $R^+(u_i)$ and $R^-(u_i)$, respectively.
All these values can be computed in linear time for both $r$-gather clustering problem and $r$-gathering problem using the same technique as in \cite{sarker2019r} (see Appendix~\ref{Ap:line} for details).

The complete algorithm is presented in Algorithm~\ref{alg} in Appendix~\ref{ap:pseudocode}.
The correctness is clear from the construction.
Thus, we analyze the time complexity.
A naive implementation of the algorithm requires $O(2^dn^2r^2d)$ evaluations of $\Cost$ and preprocessing for $R^+$ and $R^-$.
Each evaluation of $\Cost$ requires $O(1)$ time for the $r$-gather clustering problem and $O(m)$ time for the $r$-gathering problem.
The preprocessing requires $O(n)$ time for the $r$-gather clustering problem and $O(n+m)$ time for the $r$-gathering problem~\cite{sarker2019r}.
Thus, the time complexities are $O(2^dn^2r^2d)$ for the $r$-gather clustering problem and $O(2^dn^2r^2dm)$ for the $r$-gathering problem.

We can further improve the complexities of the algorithms.
The loop for $i$ is bounded to look only the first $(2r-1)d$ users from each leg because other users cannot be contained in the ball part of multi-leg clusters.
Thus, we can reduce $n$ to $rd^2$ in the complexity so as we obtain the complexities $O(2^dr^4d^5+n)$ for the $r$-gather clustering problem and $O(2^dr^4d^5m+n)$ for the $r$-gathering problem.
This proves Theorem~\ref{tractable} for the $r$-gather clustering problem.
In the $r$-gathering problem, we can further improve the complexity by improving the algorithm to calculate $\Cost$ (see Appendix~\ref{Ap:cost}).
This reduces the complexity to $O(2^dr^4d^5 + n + m)$, which proves Theorem~\ref{tractable} for the $r$-gathering problem.

\section{PTAS for r-Gathering Problem}

We prove Theorem~\ref{ptas} by demonstrating a PTAS for the $r$-gathering problem.
The technique in this section can be extended to the case that the input is a tree; see Appendix~\ref{Ap:ptas}.

As mentioned at the beginning of Section~3, the $r$-gather clustering problem is reduced to the $r$-gathering problem.
This establishes the existence of a PTAS for the $r$-gather clustering problem.


Given an instance $\mathcal{I}$ and a positive number $\epsilon > 0$, our algorithm outputs a solution whose cost is at most $(1+\epsilon)\OPT(\mathcal{I})$.
Without loss of generality, we can assume that $\epsilon\leq 1$.
First, we guess the optimal value. We can try all candidates of the optimal values because the optimal value is the distance between a user and a facility.
We then solve the corresponding (relaxed) feasibility problem whose objective value is at most the guessed optimal value.

Now we consider implementing the following oracle $\texttt{Solve}(\mathcal{I},b,\delta)$:
Given an instance $\mathcal{I}$, a threshold $b$, and a positive number $\delta$, it reports YES if $\OPT(\mathcal{I}) \le (1 + \delta) b$ and NO if $\OPT(\mathcal{I}) > b$.
If $b < \OPT(\mathcal{I}) \le (1 + \delta) b$ then both answers are acceptable.
Our oracle also outputs the corresponding solution as a certificate if it returns YES.
It should be noted that we cannot set $\delta = 0$ because it gets reduced to the decision version of the $r$-gathering problem, which is NP-hard on a spider (Theorem~\ref{hardness}).
To obtain a PTAS, we call the oracle $\texttt{Solve}(\mathcal{I},b,\epsilon)$ for each candidate of the optimal value $b$. We return the smallest $b$ that the oracle returns YES.

\subsection{Algorithm Part 1: Rounding Distance}

This and the next subsections present an implementation of $\texttt{Solve}(\mathcal{I}, b, \delta)$.
Our algorithm is a DP that maintains distance information in the indices of the DP table.
For this purpose, we round the distances so that 
the distances from the center to the vertices (thus, the users and facilities) are multiples of a positive number $t$ as follows:
For each user or facility on point $(l,x)$, we move it to the coordinate $(l,\lceil x/t \rceil)$ to make a rounded instance $\mathcal{I}'$.
Intuitively, this moves all users and facilities ``toward the center'' and regularizes the edge lengths into integers. 
Then, we define the rounded distance $d'$ on $\mathcal{I}'$.
This rounding process changes the optimal value only slightly as follows.
\begin{lemma}
For any pair of points $v$ and $w$, we have $d(v,w)\leq d'(v,w)t\leq d(v,w)+2t$.
Especially, $|\OPT(\mathcal{I})-\OPT(\mathcal{I'})t|\leq 2t$.
\end{lemma}
\begin{proof}
Let $o$ be the center of the spider.
Then, $d(v,w)=d(v,o)+d(o,w)$ holds. By definition, $d(v,o)\leq d'(v,o)t\leq d(v,o)+t$ and $d(o,w)\leq d'(o,w)t\leq d(o,w)+t$. Adding them yields the desired inequality.
\end{proof}
This lemma implies that an algorithm that determines whether $\mathcal{I}'$ has a solution whose cost is at most $b / t$ works as an oracle $\texttt{Solve}(\mathcal{I}, b, \epsilon)$ by taking $t = b \delta / 2$.

\subsection{Algorithm Part 2: Dynamic Programming}

Now we propose an algorithm to determine whether $\mathcal{I}'$ has a solution whose cost is at most $b/t$.
Because all distances between the users and the facilities of $\mathcal{I}'$ are integral, we can replace the threshold by $K = \lfloor b/t\rfloor$. 
An important observation is that $K$ is bounded by a constant because of $K\leq b/t=2/\delta$.

Now, we establish a DP. 
We define a multi-dimensional table $\S$ of boolean values such that for each integer $i$ and integer arrays $P=(p_0, \dots, p_K)$ and $Q=(q_0, \dots, q_K)$, $\S[i][P][Q]$ is $\texttt{true}$ if and only if there is a way to 
\begin{itemize}
    \item open some facilities on $l_{\leq i} := l_1\cup\dots\cup l_i$, and 
    \item assign some users on $l_{\leq i}$ to the opened facilities so that
    \begin{itemize}
    \item for all $j=0, \dots, K$, there are $p_j$ unassigned users in $l_{\leq i}$ who are distant from the center by distance $j$ and no other users are unassigned, and
    \item for all $j=0, \dots, K$, we will assign $q_j$ users out of $l_{\leq i}$ who are distant from the center by distance $j$ to the opened facilities in $l_{\leq i}$.
    \end{itemize}
\end{itemize}
Then, $\S[d][(0,\dots,0)][(0,\dots,0)]$ is the output of the $\Solve$ oracle. 
The elements of $P$ and $Q$ are non-negative integers at most $n$; thus, the size of the DP table is $O(d \times n^{2(K+1)})$, which is polynomial in the size of input.

To fill the table $\S$, we use an auxiliary boolean table $\R$ that only considers the $i$-th leg, i.e., for each integer $i$ and integer arrays $P=(p_0, \dots, p_K)$ and $Q=(q_0, \dots, q_K)$, $\R[i][P][Q]$ is $\texttt{true}$ if and only if there is a way to 
\begin{itemize}
    \item open some facilities on $l_{i}$, and
    \item assign some users on $l_{i}$ to the opened facilities so that
    \begin{itemize}
    \item for all $j=0, \dots, K$, there are $p_j$ unassigned users in $l_{i}$ who are distant from the center by distance $j$ and no other users are unassigned, and
    \item for all $j=0, \dots, K$, we will assign $q_j$ users out of $l_{i}$ who are distant from the center by distance $j$ to the opened facilities in $l_{i}$.
    \end{itemize}
\end{itemize}
We can fill the table $\R$ in polynomial time, and if we have the table $\R$, we can compute the table $\S$. Therefore we have Theorem~\ref{ptas}. Due to the space limitation, the proof is given in Appendix~\ref{Ap:ptasspider}.

\bibliographystyle{splncs04}
\bibliography{bib.bib}

\newpage

\appendix

\section{Proof of NP-Hardness of Arrears Problem}\label{Ap:hard}

We construct an instance $\mathcal{I}'$ of the arrears problem from a given instance $\mathcal{I}$ of the 1-IN-3SAT problem.
In our construction, each payment duty has exactly two payment dates.
Let us fix the variable $x_i$.
For all $j=1,\dots,m$ and $k=1,2,3$, we prepare two items $u_{i,j,k}$ and $\bar{u}_{i,j,k}$.
We also prepare auxiliary items $w_{i,l}$ and $\bar{w}_{i,l}$ for each $l=1,\dots,3m(m+1)+1$.
Accordingly, we prepare $6m(m+2)+2$ items in total for each $x_i$.

We name some important sets of items as follows.
\begin{align}
    U_i &=\{u_{i,j,k}|1\leq j\leq m,1\leq k\leq 3\}, \\
    \bar{U}_i &=\{\bar{u}_{i,j,k}|1\leq j\leq m,1\leq k\leq 3\}, \\
    W_i &=\{w_{i,l}|1\leq l\leq 3m(m+1)+1\}, \\
    \bar{W}_i &=\{\bar{w}_{i,l}|1\leq l\leq 3m(m+1)+1\}, \\
    T_i &=U_i\cup W_i, \\
    \bar{T}_i &=\bar{U}_i\cup \bar{W}_i, \\
    Y_i &=T_i\cup \bar{T}_i.
\end{align}
We prepare a payment duty $S_y=\{(a_{y,1},p_{y,1}),(a_{y,2},p_{y,2})\}$ for all items $y\in Y_i$.

We take an integer $B$ and represent all integers $p_{y,1}$ as 
\begin{align}
    p_{y,1} = e_{y,4}B^4+e_{y,3}B^3+e_{y,2}B^2+e_{y,1}B+e_{y,0}
\end{align}
for some non-negative integers $e_{y,4},\dots,e_{y,0}$.
We also represent all integers $q_j$ as 
\begin{align}
    q_j = f_{j,4}B^4+f_{j,3}B^3+f_{j,2}B^2+f_{j,1}B+f_{j,0}
\end{align}
for some non-negative integers $f_{j,4},\dots,f_{j,0}$.
By choosing $B$ sufficiently large so that any ``carry'' does not occur during the intermediate computation, we can identify these values as five-dimensional vectors 
$(e_{y,4},e_{y,3},e_{y,2},e_{y,1},e_{y,0})$ and $(f_{j,4},f_{j,3},f_{j,2},f_{j,1},f_{j,0})$ equipped with the lexicographical comparison.
Later, we wee $B = 100 n^2 m^2$ is enough.

We define the value of payment duties.
Let us fix a variable $x_i$. For $u\in U_i\cup \bar{U}_i$, we set $a_{u_1}=i$ and 
\begin{align}
a_{u,2}=\left\{\begin{array}{ll}
n+2+j & (u=u_{i,j,k}\:\text{and}\:c_{j,k}=x_i), \\
n+2+j & (u=\bar{u}_{i,j,k}\:\text{and}\:c_{j,k}=\bar{x}_i), \\
n+1 & (\text{otherwise}).
\end{array}
\right.
\end{align}
We set $p_{u,2} = 2p_{u,1}$ and 
\begin{align}
p_{u,1} = 
\left\{\begin{array}{ll}
(B^2+i)(B+1)B+(j+1) & (u=u_{i,j,k}\:\text{and}\:c_{j,k}=x_i), \\
(B^2+i)(B+1)B & (u=u_{i,j,k}\:\text{and}\:c_{j,k}\neq x_i), \\
(B^2+i)B^2+(j+1) & (u=\bar{u}_{i,j,k}\:\text{and}\:c_{j,k}=\bar{x}_i), \\
(B^2+i)B^2 & (u=\bar{u}_{i,j,k}\:\text{and}\:c_{j,k}\neq \bar{x}_i).
\end{array}
\right.
\end{align}
For each $w\in W_i\cup \bar{W}_i$,
we define $K_i = \sum_{u\in U_i}(p_{i,1} \mod B)$ and $\bar{K}_i = \sum_{u\in \bar{U}_i}(p_{i,1} \mod B)$.
We set $a_{w,1}=i$ and 
\begin{align}
a_{w,2}=\left\{\begin{array}{ll}
n+2 & (w=w_{i,l}\:\text{and}\:1\leq l\leq 3m(m+1)-K_i),\\
n+1 & (w=w_{i,l}\:\text{and}\:3m(m+1)-K_i+1\leq l\leq 3m(m+1)+1),\\
n+2 & (w=\bar{w}_{i,l}\:\text{and}\:1\leq l\leq 3m(m+1)-\bar{K}_i),\\
n+1 & (w=\bar{w}_{i,l}\:\text{and}\:3m(m+1)-\bar{K}_i+1\leq l\leq 3m(m+1)+1).
\end{array}
\right.
\end{align}
We set $p_{w,2} = 2p_{w,1}$ and
\begin{align}
& p_{w,1} = \nonumber \\
&\left\{\begin{array}{ll}
(B^2+i)(B+1)B+1 & (w=w_{i,l}\:\text{and}\:1\leq l\leq 3m(m+1)-K_i),\\
(B^2+i)(B+1)B & (w=w_{i,l}\:\text{and}\:3m(m+1)-K_i+1\leq l\leq 3m(m+1)),\\
(B^2+i)(B+1)B & (w=w_{i,l}\:\text{and}\:l=3m(m+1)+1),\\
(B^2+i)B^2+1 & (w=\bar{w}_{i,l}\:\text{and}\:1\leq l\leq 3m(m+1)-\bar{K}_i),\\
(B^2+i)B^2 & (w=\bar{w}_{i,l}\:\text{and}\:3m(m+1)-\bar{K}_i+1\leq l\leq 3m(m+1)),\\
(B^2+i)(B+3m(m+2)+1)B & (w=\bar{w}_{i,l}\:\text{and}\:l=3m(m+1)+1).
\end{array}
\right.
\end{align}
Note that, because $K_i,\bar{K}_i\leq 3m(m+1)$ by definition, 
\begin{align}
\sum_{t\in T_i}p_{t,1}=\sum_{t\in \bar{T}_i}p_{t,1}=(3m(m+2)+1)(B^2+i)(B+1)B+3m(m+1) := R
\end{align}
holds for all $i$.
We set
\begin{align}
R=\sum_{i=1}^nR_i=(3m(m+2)+1)(B+1)(nB^2+\frac{n(n+1)}{2})B+3m(m+1)n.
\end{align}
Each coefficient $e_{y,k}$ of $p_i$ is obtained by expanding the above definition.
It should be noted that the sum of $e_{y,k}$ is at most $2(3m(m+2)+1) n(n+1)/2\leq 20m^2n^2 < B$; hence, no carry is occurred.
Next, we set a budget constraint $(i,q_i)$ for each $i=1,\dots,n+m+2$. 
We set 
\begin{align}
q_i = 
\left\{\begin{array}{ll}
(3m(m+2)+1)(B+1)iB^3+(B^3-1) & (1\leq i\leq n-1),\\
R & (i=n),\\
((3m(m+2)+1)n+6mn+2n+m(m+1))B^4+(B^4-1) & (i=n+1),\\
(3(3m(m+2)+1)n-2(n+m+2-i))B^4+(B^4-1) & (n+2\leq i\leq n+m+1),\\
3R & (i=n+m+2).
\end{array}\right.
\end{align}
Each coefficient $f_{i,k}$ of $q_i$ is obtained by expanding the above definition.
This completes our construction.
All appearing values are at most $3R=O(nm^2B^4)=O(n^9m^{10})$, which is bounded in a polynomial of $n,m$.

We prove that $\mathcal{I}$ is a YES-instance of the 1-IN-3SAT problem if and only if $\mathcal{I}'$ is a YES-instance of the arrears problem.
For a feasible solution of $\mathcal{I}'$, Let $X$ be the set of items $t$ such that payment date $p_{t,2}$ is chosen for the payment duty $S_t$.
Let $\bar{X}$ be the complement of $X$.
Intuitively, for $y = u_{i,j,k}$ of $y = w_{i,l}$, 
we assign $x_i = \texttt{true}$ if $y \in X$ and $x_i = \texttt{false}$ if $y \in \bar{X}$.
The following proposition guarantees that this assignment is well-defined.
\begin{prp}\label{weldef}
In a feasible solution of $\mathcal{I}'$, for all $1\leq i\leq n$, one of the following conditions holds.
\begin{itemize}
    \item $X\cap Y_i=T_i$.
    \item $X\cap Y_i=\bar{T}_i$.
\end{itemize}
\end{prp}

Before proving this proposition, we prove the following basic property.

\begin{lemma}\label{halfeq}
In a feasible solution of $\mathcal{I}'$, 
\begin{align}
\sum_{y\in \bar{X}}p_{y,1}=\sum_{y\in X}p_{y,1}=R.
\end{align}
\end{lemma}
\begin{proof}
By the definition,
\begin{align}
\sum_{y\in \bar{X}}p_{y,1}+\sum_{y\in X}p_{y,1}=\sum_{i=1}^n\left(\sum_{y\in T_i}p_{y,1}+\sum_{y\in \bar{T}_i}p_{y,1}\right)=2R.
\end{align}
From the budget constraint for $n$, we have
\begin{align}
\sum_{y\in \bar{X}}p_{y,1}\leq R.
\end{align}
Also, from the budget constraint for $n+m+2$, we have
\begin{align}
\sum_{y\in \bar{X}}p_{y,1}=4R-\left(\sum_{y\in \bar{X}}p_{y,1}+2\sum_{y\in X}p_{y,1}\right)\geq 4R-3R=R
\end{align}
This indicates that 
bot inequalities hold in equality; thus, the lemma holds.
\end{proof}

\begin{proof}[Proof of Proposition~\ref{welldefeq}]
We consider the coefficients of $B^4$ and $B^3$.
From the budget constraint for $i=1,\dots,n$, we have
\begin{align}
\sum_{y\in \bar{X}\cap (Y_1\cup\dots\cup Y_i)}(e_{y,4},e_{y,3}) \leq (3m(m+2)+1)i(1,1)
\end{align}
for all $i=1,\dots,n-1$.
Here, $(\cdot, \cdot)$ is the inner product in the five-dimensional vector space.
Because $i(e_{y,4},e_{y,3})=(e_{y,2},e_{y,1})$ for all $y\in Y_i$,  we have
\begin{eqnarray}
(f_{n,2},f_{n,1})&=&\sum_{y\in \bar{X}}(e_{y,2},e_{y,1}) \\
&=&\sum_{i=1}^{n}\sum_{y\in \bar{X}\cap Y_i}i(e_{y,4},e_{y,3}) \\
&=&\sum_{i=1}^{n}\sum_{y\in \bar{X}\cap Y_i}n(e_{y,4},e_{y,3})-\sum_{i=1}^{n-1}\sum_{y\in \bar{X}\cap (Y_1\cup\dots\cup Y_i)}(e_{y,4},e_{y,3}) \\
&\geq&n\sum_{y\in \bar{X}}(e_{y,4},e_{y,3})-\sum_{i=1}^{n-1}(3m(m+2)+1)i(1,1)\\
&=&((3m(m+2)+1)(n^2-\frac{(n-1)n}{2})(1,1) \\
&=&(f_{n,2},f_{n,1}).
\end{eqnarray}
Therefore,
\begin{align}
\sum_{y\in \bar{X}\cap (Y_1\cup\dots\cup Y_i)}(e_{y,4},e_{y,3}) = (3m(m+2)+1)i(1,1)
\end{align}
for all $i=1,\dots,n-1$.
This equality also holds in $i = n$ because of Lemma~\ref{halfeq}.
Thus,
\begin{align}
\sum_{y\in \bar{X}\cap Y_i}(e_{y,4},e_{y,3}) = (3m(m+2)+1)(1,1)
\end{align}
holds for all $i=1,\dots,n$.
Because of the definitions of $e_{y,4}$ and $e_{y,3}$, this equation holds only if $\bar{X}\cap Y_i=T_i$ or $\bar{X}\cap Y_i=\bar{T}_i$.
\end{proof}

Next, we consider clauses.
We define the set of items $Z_j$ for all $j=0,\dots,m+1$ as the set of items $y$ with $e_{y,0}=j$.
The following proposition ensures that exactly one variable in each clause is \texttt{true}.
\begin{prp}\label{clause}
In a feasible solution of $\mathcal{I}'$, for all $j=2,\dots,m+1$, $|X\cap Z_j|=1$ holds.
\end{prp}
\begin{proof}
The proof is similar to that of Proposition~\ref{weldef}.
We consider the coefficient of $B^4$.
From the budget constraints for $n+1,\dots,n+m+1$ and Lemma~\ref{halfeq}, we have
\begin{align}
\sum_{y\in X\cap Z_0}2e_{y,4}\leq 6mn+2n+m(m+1)
\end{align}
and for all $j=1,\dots,m$, we have
\begin{align}
\sum_{y\in X\cap (Z_0\cup\dots\cup Z_j)}2e_{y,4}\leq 2(3m(m+2)+1)n-2(m+1-j).
\end{align}
Because $e_{y,0}=je_{y,4}$ holds for all $y\in Z_j$, we have
\begin{eqnarray}
f_{n+m+2,0}-f_{n,0}&=&\sum_{y\in X}2e_{y,0}\\
&=&\sum_{j=0}^{m+1}\sum_{y\in X\cap Z_j}2je_{y,4}\\
&=&\sum_{j=0}^{m+1}\sum_{y\in X\cap Z_j}2(m+1)e_{y,4}-\sum_{j=0}^{m}\sum_{y\in X\cap (Z_0\cup\dots\cup Z_j)}2e_{y,4}\\
&\geq&2(m+1)\sum_{j=0}^{m+1}\sum_{y\in X\cap Z_j}e_{y,4}-(6mn+2n+m(m+1)) \nonumber \\
&&-\sum_{j=1}^{m}(2(3m(m+2)+1)n-2(m+1-j))\\
&=&2(3m(m+2)+1)(m+1)n-(6mn+2n+m(m+1)) \nonumber \\
&&-2(3m(m+2)+1)nm+2m(m+1)-m(m+1)\\
&=&6mn(m+1)=f_{n+m+2,0}-f_{n,0}
\end{eqnarray}
Thus, for all $j=1,\dots,m$, we have
\begin{align}
\sum_{y\in X\cap (Z_0\cup\dots\cup Z_j)}2e_{y,4} = 2(3m(m+2)+1)n-2(m+1-j).
\end{align}
That implies that $|X\cap Z_j|= 1$ holds for all $j=2,\dots,m+1$.
\end{proof}

Now, we complete the reduction.
\begin{proof}[Proof of Theorem~\ref{hardness}]
We first prove that there is a polynomial-time algorithm to construct a feasible solution of $\mathcal{I}$ from a feasible solution of $\mathcal{I}'$.
For all $i=1,\dots,n$, we set $x_i$ to be \texttt{true} when $T_i\cup X=T_i$ and \texttt{false} otherwise. By the rule of construction and Propositions~\ref{weldef}, \ref{clause}, it is a feasible solution of $\mathcal{I}$.

We then prove that there is a polynomial-time algorithm to construct a feasible solution of $\mathcal{I}'$ from a feasible solution of $\mathcal{I}$.
For all $i=1,\dots,n$, we choose $T_i$ if $x_i = \texttt{true}$ and $\bar{T}_i$ otherwise.
We set $X$ as the union of all chosen sets of items.

We only have to show that this solution satisfies the budget constraints.
Because
\begin{align}
\sum_{y\in \bar{X}\cap (Y_1\cup\dots\cup Y_i)}(e_{y,4},e_{y,3})=(3m(m+2)+1)i(1,1),
\end{align}
the budget constraint for $i=1,\dots,n-1$ hold.
Because
\begin{align}
\sum_{y\in \bar{X}}p_{y,1}=R,
\end{align}
the budget constraints for $i=n$ and $i=n+m+1$ hold.
By the construction, $|X\cap Z_j|=1$ for all $j=2,\dots,m+1$. 
Therefore,
\begin{align}
\sum_{y\in \bar{X}}e_{y,4}+\sum_{y\in X\cap (Z_0\cup\dots\cup Z_j)}2e_{y,4}=3(3m(m+2)+1)n-2(m+1-j)
\end{align}
holds and the budget constraints for $i=n+2,\dots,n+m+1$ holds.
Finally,
\begin{eqnarray}
|X\cap Z_0|&=&(3m(m+2)+1)n-|X\cap Z_1|-\sum_{j=2}^{m+1}|X\cap Z_j|\\
&=&(3m(m+2)+1)n-3m(m+1)n+\sum_{j=2}^{m+1}\sum_{y\in Z_j}(e_{y,0}-1)\\
&=&3mn+n+\frac{m(m+1)}{2}
\end{eqnarray}
holds.
Thus,
\begin{align}
\sum_{y\in \bar{X}}e_{y,4}+\sum_{y\in X\cap Z_0}2e_{y,4}\leq (3m(m+2)+1)n+6mn+2n+m(m+1)
\end{align}
holds.
Therefore, the budget constraint for $i=n+1$ holds.
This completes the proof.
\end{proof}

\section{Pseudo Codes used in Section~4}
\label{ap:pseudocode}

\begin{algorithm}[H]
\caption{Suffix-Special Multi-Leg Clusters Generation\label{subalg}}    
\begin{algorithmic}[1]
\REQUIRE Set of legs $\mathcal{L}=\{l_1, \dots, l_d\}$, set of users $\mathcal{U}=\{u_1, \dots, u_n\}$, positive integer $r$
\STATE $\mathcal{C}:=\emptyset, C:=\emptyset, S:=\{1, \dots, d\}$
\FOR{$i=1, \dots, n$}
    \IF{$l(u_i) \in S$}
        \STATE Choose one. (a): Use $u_i$ in ball part of cluster $C$. $C:=C\cup\{u_i\}$.\\ (b): Discard $u_i$ and all further users in $l(u_i)$. $S:=S\setminus \{l(u_i)\}$. In this case, discarded users are used in single-leg clusters.
    \ENDIF
    \STATE Choose one. (c): Continue to choose ball part. Do nothing. We can choose this alternative only when $|C|<2r-1$. \\ (d): Finish to choose the ball part and go on to choose the segment part.
    \IF{(d) is chosen}
        \STATE Choose a leg $l\in S$.
        \STATE Choose a non-negative integer $t$, subject to there are at least $t$ users on leg $l$ whose indices are larger than $i$ and $r\leq |C|+t\leq 2r-1$. Add first $t$ users among such users to $C$.
        \STATE $S:=S\setminus \{l\}$, $\mathcal{C}:=\mathcal{C}\cup \{C\}, C:=\emptyset$
    \ENDIF
\ENDFOR
\RETURN $\mathcal{C}$ (only when $C$ is empty)
\end{algorithmic}
\end{algorithm}

\begin{algorithm}[H]
\caption{A FPT algorithm for $r$-gather clustering problem and $r$-gathering problem on spider}         
\label{alg}             
\begin{algorithmic}[1]
\REQUIRE Set of legs $\mathcal{L}=\{l_1, \dots, l_d\}$, set of users $\mathcal{U}=\{u_1, \dots, u_n\}$, positive integer $r$. In $r$-gathering, we are also given a set of facilities $\mathcal{F}$.
\STATE Calculate $R^+(u_i),R^-(u_i)$ for all $i$.
\STATE $\DP[i][S][j][k]:= \infty$ for all $0\leq i\leq n, S\subseteq \{1,\dots d\}, 0\leq j\leq 2r, 0\leq k\leq 2dr$
\COMMENT{Here $i,S,j,k$ means the user we are looking at now, the set of available legs, the size of current cluster, the last user in current cluster, respectively}
\STATE $\DP[0][S][0][0]:=0$
\FOR{$i=1,\dots,n$}
    \FOR{$S\subseteq \{1,\dots,d\}, j=0,\dots,2r-2, k=0,\dots,i-1$ such that $l(u_i)\in S$}
        \STATE $\DP[i][S][j+1][i]:=\min(\DP[i][S][j+1][i],\DP[i-1][S][j][k])$
        \COMMENT{Here we use $u_i$ in the ball part of current cluster}
        \STATE $\DP[i][S\setminus \{l(u_{i})\}][j][k]:=\min(\DP[i][S\setminus \{l(u_{i})\}][j][k],\max(\DP[i-1][S][j][k],R^-(u_i)))$
        \COMMENT{Here we discard leg $l(u_i)$}
    \ENDFOR
    \FOR{$S\subseteq \{1,\dots,d\}, j=0,\dots,2r-2, k=1,\dots,i-1, l=1,\dots,d$ such that $l\not \in S$}
         \FOR{$p=\max(0,r-j),\dots,2r-1-j$}
             \IF{There are at least $p$ users on leg $l$ whose indices are larger than $i$}
             \STATE Let $v$ be the $p$-th such user
             \STATE $\DP[i][S\setminus \{l\}][0][i]=\min(\DP[i][S\setminus\{l\}][0][i],$ $\max(\DP[i][S][j][k], \Cost(v,u_k), R^+(v)))$
             \COMMENT{Here we use remaining first $p$ users on leg $l$ as the segment part of the current cluster}
             \ENDIF
         \ENDFOR
    \ENDFOR
\ENDFOR
\RETURN{$\min \{ \DP[i][\emptyset][0][i] : i = 0, \dots, n \}$}
\end{algorithmic}
\end{algorithm}

\section{Linear-time Algorithm for r-gathering on line}\label{Ap:line}

We explain a linear-time algorithm for the $r$-gathering problem, which is given in~\cite{sarker2019r}.
This is used in our proposed algorithm as pre-processing.
We use the following lemma.
\begin{lemma}[{\rm{Lemma 1,~\cite{nakano2018simple}}}]
There is an optimal solution that the users assigned to the same facility are consecutive.
\end{lemma}
Let $\DP[i]$ be the maximum distance between a user and the assigned facility when we assign users $u_1,\dots,u_i$.
Let $\Cost(i,j)$ be the minimum cost to assign users $u_i,\dots,u_j$ to the same facility.
Then, $\DP[0]=0$ and
\begin{align}
\DP[i]=\min_{0\leq j\leq i-r}\max(DP[j],\Cost(i+1,j))
\end{align}
holds.
Thus, we obtain an algorithm of $O(n^2)$ time with $O(n^2)$ calls of $\Cost$ oracle.

\cite{sarker2019r} shows that the time complexity to calculate $\DP[i]$ is amortized $O(1)$ with $O(1)$ calls of $\Cost$ oracle using the sliding window technique.
They also constructed an algorithm to calculate all $\Cost$ values appearing in the algorithm in $O(m)$ time in total, therefore this is the $O(n+m)$ time algorithm.

For the $r$-gather clustering problem, we obtain $\Cost(i,j)$ is obtained by taking the distance between $i$ and $j$; thus, computed in $O(1)$ time.

\section{Calculation of Cost in r-gathering}\label{Ap:cost}

We show how to calculate the $\Cost(v,u)$ in the $r$-gathering problem efficiently.
The number of candidates of pair $v,u$ is at most $O(r^2d^4)$; therefore, we calculate the $\Cost$ for all candidates in advance and store them.
Now, we describe how to calculate these values.
We assume that the facilities are given in increasing order of the distances from the center.

There are two cases of the location of a facility, which will be assigned to the cluster -- located on the leg $l(v)$ or not.
If it is not located on the leg $l(v)$, we choose the facility that is closest to the center.
This case can be processed in $O(1)$ time for each pair of $v,u$.
If it is located on the leg $l(v)$, we choose the facility that is closest to the midpoint of the coordinates of $v$ and $u$.
By calculating the midpoints of all pairs and sorting them by the distance from the center for each leg with the help of the two-pointer technique, we obtain an optimal facility in $O(r^2d^4\log(rd)+m)$ time. 

We can retrieve each pre-calculated $\Cost$ value in $O(1)$ time.
Thus, the total time complexity of Algorithm \ref{alg} is reduced to $O(2^dr^4d^5+r^2d^4\log(rd)+n+m)=O(2^dr^4d^5+n+m)$.

\section{DP Transitions of PTAS on Spider}\label{Ap:ptasspider}

Here we describe the remaining part of the PTAS on a spider, that is, the ways to fill the tables $\S$ and $\R$.

If we have the table $\R$, we can easily compute the table $\S$ as follows:
For arrays $X$ and $Y$, we denote by $X + Y$ and $X - Y$ the the element-wise addition and subtraction, respectively. 
If $\S[i-1][P][Q]$ is $\texttt{true}$, for all arrays $P_1,Q_1,P_2,Q_2$ of length $K+1$ such that $\R[i][P_1+P_2][Q_1+Q_2]$ is $\texttt{true}$, $\S[i][P-Q_1+P_2][Q-P_1+Q_2]$ is also $\texttt{true}$.
Here, the $k$-th ($0$-origin) elements of $P_1$ (resp. $P_2$) represents the number of users on leg $l_i$, which are distant from the center by $k$, and assigned to the facilities on $l_{\leq i-1}$ (resp. out of $l_{\leq i}$).
Similarly, the $k$-th element of $Q_1$ (resp. $Q_2$) represents the number of users assigned to the facilities on leg $l_i$ and distant from the center by $k$, which is located on $l_{\leq i-1}$ (resp. out of $l_{\leq i}$).
All entries of $\S$ not filled by the above procedure are $\texttt{false}$.
The correctness of this DP is clear from the definitions of $\S$ and $\R$.

The remaining task is to fill the table $\R$.
Fix a leg $l_i$.
To compute $\R[i][*][*]$, we use another auxiliary boolean table $\DP$.
Let $t_1, \dots, t_k$ be the users and facilities on $l_i$ in the descending order of the distance, denoted by $x(t_i)$, from the center.
Let $U_{\leq i}$ and $F_{\leq i}$ be the users and facilities in $\{t_1,\dots,t_i\}$, respectively.
For an integer $0\leq i\leq k$ and integer arrays $P=(p_0, \dots, p_K)$ and $Q=(q_0, \dots, q_K)$, $\DP[i][P][Q]$ is $\texttt{true}$ if and only if there is a way to
\begin{itemize}
    \item open some facilities on $F_{\leq i}$, and
    \item assign some users on $U_{\leq i}$ to the opened facilities so that
    \begin{itemize}
    \item for all $j=0, \dots, K$, there are $p_j$ unassigned users in $U_{\leq i}$ who are distant from $t_i$ by distance $j$ and no other users are unassigned, and
    \item for all $j=0, \dots, K$, we will assign $q_j$ users out of $U_{\leq i}$ who are distant from $t_i$ by distance $j$ to the opened facilities in $F_{\leq i}$.
    \end{itemize}
\end{itemize}
Then, $\R[i][P][Q] = \DP[k][P^{-x(t_k)}][Q^{x(t_k)}]$ holds, where $P^k$ is the the array produced by shifting $P$ by $k$ rightwards if $k \ge 0$ and the array produced by shifting $P$ by $|k|$ leftwards if $k < 0$; the overflowed entries are discarded.

Next, we present an algorithm to calculate the $\DP$ table. 
The transitions are as follows:
If $t_i$ is a user and $\DP[i-1][P][Q]$ is $\texttt{true}$, then 
$\DP[i][P^{(x(t_i)-x(t_{i-1}))}+(1,0,\dots,0)][Q^{-(x(t_i)-x(t_{i-1}))}]$ and $\DP[i][P^{(x(t_i)-x(t_{i-1}))}][Q^{-(x(t_i)-x(t_{i-1}))}-(1,0,\dots,0)]$
are also $\texttt{true}$.
The first transition assigns $t_i$ to an already opened facility, and the second transition assigns $t_i$ to a facility that will be opened in future. 
All entries of $\DP[i][*][*]$ not filled by the above transition are $\texttt{false}$.
If $t_i$ is a facility and $\DP[i-1][P][Q]$ is $\texttt{true}$, then $\DP[i][P^{(x(t_i)-x(t_{i-1}))}-P'][Q^{-(x(t_i)-x(t_{i-1}))}+Q']$ is also $\texttt{true}$ 
for all integer arrays $P'$ and $Q'$ satisfying that
\begin{itemize}
    \item
    the lengths of $P',Q'$ are both $K+1$,
    \item 
    the sums of entries of $P'$ and $Q'$ are at least $r$, and
    \item 
    the sums of indices of the last non-zero elements of $P'$ and $Q'$ ($0$-origin) are at most $r$.
\end{itemize}
Here, $P'$ (resp. $Q'$) represents the list of the distances of the users assigned to $t_i$, which is before (resp. after) $t_i$.
All the states with form $\DP[i][P'][Q']$ which cannot be represented in above formula are $\texttt{false}$.

We can reconstruct the solution by storing 
the transition candidates that were chosen.
Thus, we constructed the desired algorithm, which proves Theorem~\ref{ptas}.

\section{PTAS on Tree}\label{Ap:ptas}

We extend the result of Section~5 to a tree, i.e., we prove the following theorem.
\begin{theorem}
\label{ptas-tree}
There are PTASes to the $r$-gather clustering problem and $r$-gathering problem on a tree.
\end{theorem}

A \emph{weighted tree} $T = (V(T), E(T); l)$ is an undirected connected graph without cycles, where $V(T)$ is the set of vertices, $E(T)$ is the set of edges, and $l \colon E(T) \to \mathbb{R}_+$ is the non-negative edge length.
$T$ forms a metric space by the tree metric $d(v, w)$, which is the sum of the edge lengths on the unique simple $v$-$w$ path for any vertices $v, w \in V(T)$.

We propose an algorithm to the $r$-gathering problem on this metric space. 
Because the $r$-gather clustering problem is a special case of the $r$-gathering problem, the algorithm for the $r$-gathering problem can also be applied to the $r$-gather clustering problem in a straightforward way.

Without loss of generality, we assume that all users and facilities are located on different vertices; otherwise, we add new vertices connected with edges of length zero and separate the users and facilities into the new vertices. 
We also assume that $T$ is a rooted full binary tree rooted at a special vertex $\root$ (i.e., each vertex has zero or two children).
This only increases the number of vertices (and edges) by a constant factor; thus, this does not affect the time complexity of our algorithms.
We denote the subtree of $T$ rooted at $v$ by $T_v$.

As same as the spider case, the proposed algorithm consists of the $\Solve$ oracle.
Also, as same as the spider case, $\Solve$ oracle consists of the distance rounding and DP.
Both parts are similar to those of the spider case; however, to handle the tree structure, the details become much complicated, especially on the DP part.

\subsection{Rounding Distance}

For each edge $e = (v, w) \in E(T)$, where $v$ is closer to the root, we define the rounded length by $l'(e) = \lfloor d(\root,w)/t \rfloor-\lfloor d(\root,v)/t\rfloor$.
Then, we define the rounded distance $d'$ the metric on $\mathcal{I}'$.
This rounding process changes the optimal value only slightly as follows.
\begin{lemma}
For any pair of vertices $v, w$, $d(v,w)-2t\leq d'(v,w)t\leq d(v,w)+2t$ holds.
Especially, $|\OPT(\mathcal{I})-\OPT(\mathcal{I'})t|\leq 2t$.
\end{lemma}
\begin{proof}
Let $x$ be the lowest common ancestor of $v$ and $w$.
Then, $x$ is on the $v$-$w$ path; thus, $d(v,w) = d(x,v) + d(x,w)$ and $d'(v,w)=d'(x,v)+d'(x,w)$ hold.
Because $d(x,v)=d(\root,v)-d(\root,x), d'(x,v)=d'(\root,v)-d'(\root,x)$ and $d(\root,z)-t\leq d'(\root,z)t\leq d(\root,z)$ for all vertex $z$, we have $d(x,v)-t\leq d'(x,v)t\leq d(x,v)+t$. 
We also have $d(x,w)-t\leq d'(x,w)t\leq d(x,w)+t$ by symmetry.
Thus $d(v,w)-2t\leq d'(v,w)t\leq d(v,w)+2t$ holds.
Because the cost of the $r$-gathering problem is the maximum length of some paths, the second statement follows from the first statement.
\end{proof}
This lemma implies that any algorithm that determines whether $\mathcal{I}'$ has a solution with cost at most $(b+2t)/t$ can be used as an oracle $\texttt{Solve}(\mathcal{I}, b, \epsilon)$ if $t = b \delta/4$.

\subsection{Dynamic Programming}

Now, we propose an algorithm to determine whether $\mathcal{I}'$ has a solution with cost at most $(b+2t)/t$.
Because all the edge costs of $\mathcal{I}'$ are integral, without loss of generality, we replace the threshold by $K = \lfloor (b+2t)/t \rfloor$.
It should be noted that $K$ is bounded by a constant because $K\leq (b+2t)/t=(4/\delta) + 2$.

Our algorithm is a DP on a tree.
For vertex $v$ and integer arrays $P=(p_0, \dots, p_K)$ and $Q=(q_0, \dots, q_K)$, we define a boolean value $\DP[v][P][Q]$.
$\DP[v][P][Q]$ is $\texttt{true}$ if and only if there is a way to 
\begin{itemize}
    \item open some facilities in $T_v$, and
    \item assign some users in $T_v$ to the opened facilities so that
    \begin{itemize}
    \item for all $i=0, \dots, K$, there are $p_i$ unassigned users in $T_v$ who are distant from $v$ by distance $i$ and no other users are unassigned, and
    \item for all $i=0, \dots, K$, we will assign $q_i$ users out of $T_v$ who are distant from $v$ by distance $i$ to the opened facilities in $T_v$,
    \end{itemize}
\end{itemize}
Then, $\DP[\root][(0, \dots, 0)][(0,\dots, 0)]$ is the solution to the oracle.
The elements of $P$ and $Q$ are non-negative integers at most $n$; thus, the number of the DP states is $|V(T)| \times (n+1)^{2(K+1)}$, which remains in polynomial in the size of input.

We define the transition of the DP.
Let $x, y$ be the two children of $v$ and $d_x, d_y$ be the cost of edges $(v,x),(v,y)$, respectively.
Then, $\DP[v][P(v)][Q(v)]$ is $\texttt{true}$ if and only if
\begin{itemize}
    \item there are arrays $P(x),Q(x),P(y),Q(y),R(x),R(y),S_1,S_2,W_1,W_2$ of integers whose lengths are $K+1$ such that 
    \item $S_1+S_2$ is $(1,0, \dots, 0)$ if there is a user on $v$ and $(0, \dots, 0)$ otherwise, and 
    \item the sum of all elements in $W_1+W_2$ is zero or at least $r$ if there is a facility on $v$ and zero otherwise, and
    \item if $W_1+W_2$ is nonzero, the sum of indices of last nonzero elements of $W_1$ and $W_2$ are at most $K$, and
    \item $R(x)\leq P(x)^{d_x},Q(y)^{d_y}$ and $R(y)^{d_y}\leq P(y)^{d_y},Q(x)^{d_x}$, and
    \item $\DP[x][P(x)][Q(x)]=\DP[y][P(y)][Q(y)]=\texttt{true}$, and 
    \item $p(x)_i=0$ for $i>K-d_x$, $q(x)_i=0$ for $i<d_x$, $p(y)_i=0$ for $i>K-d_y$, $q(y)_i=0$ for $i<d_y$, and
    \item $P(v)=P(x)^{d_x}+P(y)^{d_y}-R(x)-R(y)+S_1-S_2-W_1$, and
    \item $Q(v)=Q(x)^{-d_x}+Q(y)^{-d_y}-R(x)-R(y)+W_2$.
\end{itemize}
The meanings of the auxiliary variables $R(x), R(y), S_1, S_2, W_1, W_2$ are the following:
\begin{itemize}
    \item The $i$-th entry of $R(x)$ (resp. $R(y)$) denotes the number of users in $T_x$ (resp. $T_y$) who are distant from $v$ by distance $i$ and assigned to the facility in $T_y$ (resp. $T_x$).
    \item $S_1$ and $S_2$ decide whether we assign the user on $v$ to an open facility in $T_v$ or remain unassigned.
    \item The $i$-th entry of $W_{1}$ (resp. $W_2$) denotes the number of users in $T_v$ (resp. outside of $T_v$) who are assigned to the facility on $v$ and distant from $v$ by distance $i$.
\end{itemize}
We can enumerate all possibilities of the arrays in polynomial-time.
Thus, the total time complexity is polynomial.

We can reconstruct the solution by storing the transition candidates that were chosen. 
Thus, we constructed the desired algorithm.
This gives a proof of Theorem~\ref{ptas-tree}.
\end{document}